# Sliding charge density waves and zero-resistance states in GaAs/AlGaAs heterostructures

J. C. Phillips

Dept. of Physics and Astronomy,

Rutgers University, Piscataway, N. J., 08854-8019

**Mani *et al.* have observed[1] zero-resistance states (ZRS) and energy gaps in a surprising setting: ultrahigh-mobility GaAs/AlGaAs heterostructures that contain a two dimensional electron Landau system (2DELS) exhibit vanishing diagonal resistance without Hall resistance quantization at low temperatures and low magnetic fields when the specimen is subjected to electromagnetic wave excitation. Zero-resistance-states occur about magnetic fields $B = 4/5\ B_f$ and $B = 4/9\ B_f$, where $B_f = 2\pi f m^*/e$, $m^*$ is the electron mass, $e$ is the electron charge, and $f$ is the electromagnetic-wave frequency. Activated transport measurements on the resistance minima also indicate an energy gap at the Fermi level. The results suggest an unexpected radiation-induced, electronic-state-transition in the GaAs/AlGaAs 2DELS. The authors have explained the general features of their data as possibly arising from superconductivity through exciton exchange. Here we show that there is another plausible mechanism based on self-organized sliding charge density waves that does not require superconductivity to exist in the presence of a very strong magnetic field.**

Mani *et al*. have shown that the microwave field drastically affects the resistivity of their 2DELS (not surprising), with a crossover from Shubnikov–deHaas oscillations to new low-field oscillations near the cyclotron frequency. The oscillations[2] include wide valleys of zero resistivity (very surprising), with a '1/4-cycle phase shift' of the extrema with respect to integral cyclotron frequencies. They explain the zero resistivity and the phase shift by a construction that involves a surplus energy (energy associated with confined z-axis motion) model (their Fig. 4), from which they conclude that the surplus energy has been used to form an energy gap, probably superconductive. There is a different mechanism that reflects the glassy disorder of the remote dopants and the clustering of the carriers to form domains, which seems to explain the data equally well, and does not require superconductivity to exist in the presence of a very strong magnetic field. This mechanism suggests a scenario regarding the relaxation of the resistance after the radiation has been turned off.

The basic idea is that in the presence of the microwave field for nearly ideally •-doped samples the lateral (x.y) fluctuations in the potential landscape of the depletion layer[3] lift the very high levels of degeneracy normally assumed to be present in Landau levels because of lateral translational invariance. These degeneracies are in fact lifted even in the absence of the microwave field, but in order to optimize the conditions for observing Quantum Hall Effects (QHE), the glassy system is frozen into a many-electron ground state that minimizes scattering from lateral (x.y) potential fluctuations. This effectively restores translational invariance throughout the sample (except at the edges, where skipping states are formed), and leads to the usual QHE oscillations at integral filling. The microwave field has dimensions larger than those of the sample, so that it is not equivalent merely to a thermal bath. The external field enables the many-electron system to access an exponentially large number but still topologically distinctive array of excited states. Locally these can be visualized as combinations of cyclotron orbits[2] of slightly variable radius with slightly variable centers of density along the z axis normal to the depletion layer. In other words, the 2DELS is now represented in configuration space by a dynamically fluctuating array of inequivalent pancakes in the potential landscape *organized* by the long-wavelength applied microwave field. Globally these field-induced states can be regarded as optimal-screening filaments tracing the directions of the microwave field[4].

How will such a many-electron glass respond to a dc applied electric field? An obvious possibility is that current will flow via quantum percolation along self-organized paths created by the microwave field and aligned by the dc field. (One says that the latter are self-organized, because the system can lower its free energy by optimally screening first the microwave field and then the dc applied field.) One has, in effect, a sliding charge density wave (SCDW) moving along a curvilinear path in many-electron configuration space. In fact, such a scenario is not merely speculative, because parts of the scenario have recently been observed[5] in studies of SCDW in $NbSe_3$, which has a chain-like structure especially favorable for SCDW. (The reader, of course, recalls that at rest CDW's are pinned by impurities. Above a threshold field $F_c$ (or current $I_c$), depinning occurs, and the CDW slides almost freely (nearly zero differential resistivity from the part of electron system that has formed the CDW)[6].) In very high quality samples[5] two effects appear: longitudinal ordering (longer longitudinal correlation length for $I/I_c > 1.5$), and transverse ordering (channel narrowing or reduced zigzaging) for $I/I_c > 2.5-3$. There is some hysteresis, but the relaxation has not been reported in detail.

Several models have discussed the theoretical aspects of motional ordering of SCDW[5]. The simplest idea is a single-particle one: at large velocities there is motional narrowing of the interactions with impurities. A more sophisticated approach, and more relevant to our present approach to Landau systems, is topological, and predicts two transitions, a plastic flow at intermediate fields, and a dynamic one at larger fields; both of these appear to have been observed in the transverse ordering. The noise spectrum of the SCDW is also best understood topologically in terms of weakly interacting channels[5].

Returning to the 2DELS, one sees that there is one very favorable feature that differentiates it from the $NbSe_3$ case: in the latter, there is always some background resistance associated with parts of the Fermi surface that do not contain a Peierls CDW gap, so zero SCDW resistivity cannot be observed. In the 2DELS the only current-carrying parts of the system are the self-organized quantum percolation paths, and so long as these form non-interacting channels, analogous to single-mode optical fibers, the conditions are ideal for ZRS. Of course, the formation of such specific low-entropy states requires surplus configurational free energy, and it is the latter that accounts for the observed periodicities[1].

There is one other analogy that warrants discussion, and that is the one with high-temperature superconductivity in the cuprates, or more generally, the intermediate topological filamentary phase that has been postulated not only for the cuprates, but also for many network glasses as well[7-9]. Both motionally ordered SCDW in crystals and ZRS states in 2DELS fit in very nicely with the filamentary model for HTSC; in particular, these two examples show that while superconductivity can enhance filamentary self-organization, it is not necessary to achieve it. Thus filaments can be formed at high temperatures (as high as 700K in the cuprates) by self-organization of the dopant configuration, and this accounts for the normal-state transport anomalies and pseudogap formation[7].

The fact that there are ZRS in microwave-activated 2DELS, and that the overall differences in the resistivity in the presence and absence of the microwave field are so large, presents science with an unparalleled opportunity to study configurational relaxation of a quantum glass. It would be rash of theory to even attempt to predict what the transport properties of the 2DELS would be as the microwave field is turned off and it returns to its ground state. However, the reader may consider the following analogy. No matter how large the many-particle configuration space, the motion of the system may still be diffusive, with each filament diffusing eventually to a "trap", where it breaks and ceases to contribute to the conductivity. Over time such diffusive depletion of the filamentary population takes place more and more slowly, as the filaments closest to sinks have disappeared, and only those furthest away remain. This is the Scher-Lax-Phillips (SLP) model of relaxation of persistent conductivities in amorphous semiconductors and structural relaxation of glasses. It predicts relaxation described by the real-time stretched exponential $\exp(-(t/\bullet)^{\bullet})$ in microscopically disordered but otherwise homogeneous glasses. Intrinsic stretching parameters • in many off-lattice electronic and molecular glasses exhibit a characteristic duality at two pure fractions, 3/5 and 3/7, predicted by theory[10]. Stretched exponential relaxation (SER) accompanied by the *predicted* • duality at 3/5 and 3/7 has been observed both for conductivity and hydrostatic stress relaxation in cuprates[11]. The results for • are very similar to what has been observed for picosecond carrier relaxation in semiconductor nanocrystallites (fullerene, Cd(S,Se))[10]. Should SER be observed in 2DELS, one would be able to examine •(H) throughout the entire range of reordering and to test the validity of the SLP model. A small caveat: there are many more sophisticated relaxation patterns possible. For instance, as the field is turned off, the filaments could revert from narrow single-mode dynamical channels, to broader, multi-mode branched or plastically deformed channels[4]. Or the act of switching off the field may be in itself destructive. However, the lifetimes $\bullet_{SdH}$ = 2.5 ps, $\bullet_{CR}$ = 13 ps, and $\bullet_t$ = •m*/e = 115 ps, where • is the mobility, reported in the earlier experiment[2] are consistent with diffusion, as a simple model based on projection operators[12] predicts $\bullet_{SdH} \bullet_t = \bullet_{CR}^2$. (The values in ref.1 are about 5x larger.)

Self-organization is the key to understanding semiclassical, low-field ZRS. Ordinarily, one uses this phrase in biophysical contexts, but it is also useful in some disordered systems. Note that in the biophysical context there are generally two kinds of functional scaling. Most biological functions are performed by structures that scale geometrically with mass (such as cortical functions in primates)[13], but metabolic networks exhibit optimized allometric scaling involving surface/volume ratios in spaces of higher dimensionality d* = d + 1 (warm-blooded animals) or 2d (cold-blooded)[14]. The higher dimensionality is the signature of self-organization. Here the metabolic role is played by the microwave field, as it provides a steady supply of low-entropy (long wave length) free energy that is utilized by the sample to form self-organized quantum percolation paths for SCDW.